\documentclass{aa}
\usepackage{txfonts}
\usepackage{graphicx}
\usepackage{subcaption}
\usepackage[colorlinks=true,allcolors=blue]{hyperref}
\usepackage{natbib}

\begin{document} 

\titlerunning{Comparative analysis of the SFR of AGN and non-AGN galaxies}
\authorrunning{Mountrichase et al. }
\titlerunning{Comparative analysis of the SFR of AGN and non-AGN galaxies}

\title{Comparative analysis of the SFR of AGN and non-AGN galaxies, as a function of stellar mass, AGN power, cosmic time and obscuration}

\author{G. Mountrichas\inst{1}, V. A. Masoura\inst{1}, A. Corral\inst{1} and F. J. Carrera\inst{1}}
          
    \institute {Instituto de Fisica de Cantabria (CSIC-Universidad de Cantabria), Avenida de los Castros, 39005 Santander, Spain
              \email{gmountrichas@gmail.com}
}

\abstract{This study involves a comparative analysis of the star-formation rates (SFRs) of AGN and non-AGN galaxies and of the SFRs of type 1 and 2 AGN. To carry out this investigation, we assembled a dataset consisting of 2\,677 X-ray AGN detected by the XMM-Newton observatory and a control sample of 64\,556 galaxies devoid of AGN. We generated spectral energy distributions (SEDs) for these objects using photometric data from the DES, VHS, and AllWISE surveys, and we harnessed the CIGALE code to extract measurements for the (host) galaxy properties. Our dataset encompasses a diverse parameter space, with objects spanning a range of stellar masses from $\rm 9.5<\log\,[M_*(M_\odot)]<12.0$, intrinsic X-ray luminosities within $\rm 42<\log\,[L_{X,2-10keV}(ergs^{-1})]<45.5$, and redshifts between $\rm 0.3<z<2.5$. To compare SFRs, we calculated the SFR$_{norm}$ parameter, which signifies the ratio of the SFR of AGN to the SFR of non-AGN galaxies sharing similar M$_*$ and redshift. Our analysis revealed that systems hosting AGN tend to exhibit elevated SFRs compared to non-AGN galaxies, particularly beyond a certain threshold in L$_X$. Notably, this threshold increases as we move towards more massive galaxies. Additionally, for AGN systems with the same L$_X$, the magnitude of the SFR$_{norm}$ decreases as we consider more massive galaxies. This suggests that in galaxies with AGN, the increase in SFR as a function of stellar mass is not as prominent as in galaxies without AGN. This interpretation finds support in the shallower slope we identify in the X-ray star-forming main sequence in contrast to the galaxy main sequence. Employing CIGALE's measurements, we classified AGN into type 1 and type 2. In our investigation, we focused on a subset of 652 type 1 AGN and 293 type 2 AGN within the stellar mass range of $\rm 10.5<\log,[M_(M_\odot)]<11.5$. Based on our results, type 1 AGN display higher SFRs than type 2 AGN, at redshifts below $\rm z<1$. However, at higher redshifts, the SFRs of the two AGN populations tend to be similar. At redshifts $\rm z<1$, type 1 AGN show augmented SFRs in comparison to non-AGN galaxies. In contrast, type 2 AGN exhibit lower SFRs when compared to galaxies that do not host an AGN, at least up to $\rm log\,[L_{X,2-10keV}(ergs^{-1})]<45$.}

   
\maketitle

\section{Introduction}

Most, if not all, galaxies host a supermassive black hole (SMBH) in their centre. These SMBHs become active when material is accreted onto them. This process produces copious amounts of energy that can be observed as intense radiation at different wavelengths (X-ray, UV, mid-infrared, radio) and constitutes the characteristic signature of the class of Active Galactic Nuclei (AGN). The energy released during the accretion process is also an important source of heating for both the interstellar and intergalactic medium \citep[e.g.,][] {Morganti2017}. As a result, it has been hypothesized that AGN activity plays an important role in both galaxy evolution and more generally structure formation in the Universe \citep[e.g.][]{Brandt_Alexander2015}. However, establishing such a connection necessitates addressing critical questions, including the existence of a correlation between AGN activity and baryonic phenomena such as star formation. Moreover, it has been shown that most of the energy emitted by radiation in the universe is obscured \citep[e.g.,][]{Akylas2006}. Thus, another crucial aspect of this undertaking is to uncover the physical differences that distinguish between obscured and unobscured AGN.

Numerous studies have tried to tackle the first question of whether a relationship exists between AGN activity, using as a proxy the X-ray luminosity (L$_X$), and star formation \citep[e.g.,][]{Lutz2010, Page2012,Harrison2012, Rosario2012, Santini2012, Rovilos2012, Shimizu2015, Rosario2013, Mullaney2015, Masoura2018, Bernhard2019, Florez2020, Torbaniuk2021, Torbaniuk2023}. However, the outcomes of these investigations are conflicting. Some studies have identified that galaxies with low-to-moderate L$_X$ (L$\rm _X<10^{43.5}\,erg\,s^{-1}$) exhibit enhanced star formation compared to non-AGN galaxies, and this trend becomes more pronounced at higher L$_X$ \citep{Santini2012}. Conversely, other research has found similar star formation rates (SFR) between the two populations for low-to-moderate L$_X$ AGN \citep{Bernhard2019}, and a reduced SFR in luminous AGN compared to non-AGN systems \citep{Shimizu2015}. Additionally, a more intricate relationship between the two properties (SFR, L$_X$), contingent on the AGN's position relative to the star-formation main-sequence \citep[MS; e.g.,][]{Noeske2007, Elbaz2007, Whitaker2012, Speagle2014}, has been reported as well \citep{Masoura2018}.

More recently, \cite{Mountrichas2021b, Mountrichas2022a, Mountrichas2022b} conducted a comprehensive analysis by comparing the SFR of AGN with that of non-AGN galaxies. They considered a wide range of X-ray luminosities ($\rm 42.5<log\,[L_{X,2-10keV}(ergs^{-1})]<44.5$) and redshifts ($\rm 0.0<z<2.5$) using data from the Bo$\rm \ddot{o}$tes, COSMOS and eFEDS field. To facilitate this investigation, they compiled a reference galaxy catalog that shared the same photometric coverage as the X-ray sources. The research involved constructing and fitting spectral energy distributions (SEDs) for both the X-ray and galaxy samples using identical modules and parametric grids. This approach aimed to minimize systematic effects in the analysis. Their results showed that AGN with intermediate stellar mass ($\rm 10.5<log\,[M_*(M_\odot)]<11.5$) tend to have lower or at most equal SFR compared to galaxies without AGN at low-to-moderate L$_X$ ($\rm log\,[L_{X,2-10keV}(ergs^{-1})]<44$). However, more luminous X-ray sources demonstrated enhanced SFR (by $\sim 30\%$) compared to non-AGN galaxies. One of the limitations of these studies, though, was the small number of X-ray sources that probed low M$_*$ (i.e., $\rm log\,[M_*(M_\odot)]<10.5$) and very high luminosities ($\rm log\,[L_{X,2-10keV}(ergs^{-1})]>44.5$).

Regarding AGN obscuration, two primary models aim to elucidate the underlying mechanisms. The unification model \citep[e.g.][]{Urry1995, Nenkova2002, Netzer2015} classifies AGN based on the observer's line of sight relative to the central black hole's accretion disk. Obscured AGN are seen edge-on, unobscured face-on. Evolutionary models suggest that different AGN types result from SMBH and host galaxy evolutionary phases. Obscured AGN, seen in an early phase, lack the energy to disperse surrounding gas. As material accumulates, energy intensifies, causing the obscuring material to dissipate \citep[e.g.,][]{Ciotti1997, Hopkins2006}. 

Previous studies, which employed optical criteria like optical spectra to categorize X-ray AGN into type 1 and type 2, have observed that type 2 sources typically reside in more massive systems compared to type 1. However, they did not find statistically significant distinctions in the star-formation rate (SFR) between these two AGN populations \citep[e.g.,][]{Zou2019, Mountrichas2021b}. In a recent study by \cite{Mountrichas2024a}, X-ray AGN data from the eFEDS and COSMOS fields were analyzed. This research confirmed the previous findings regarding the M$_*$ differences between the two AGN populations. However, their analysis unveiled variations in the SFR of type 1 and type 2 AGN that were contingent on the L$_X$ and redshift of the sources. Specifically, it was observed that type 2 sources exhibited lower SFRs compared to type 1 AGN at $\rm z<1$. Interestingly, this trend reversed for sources at $\rm z>2$ and with high L$_X$ ($\rm log\,[L_{X,2-10keV}(ergs^{-1})]>44$).

In this study, we employ X-ray sources detected by the XMM-Newton observatory and compile a control sample of non-AGN galaxies using data from the DES, VHS, and AllWISE surveys, within the XMM footprint. We then generate spectral energy distributions (SEDs) for both galaxy populations and utilize SED fitting techniques with the CIGALE code. Furthermore, we make use of CIGALE's measurements to categorize sources into type 1 and type 2 AGN. Our study is driven by two primary objectives.  Firstly, we endeavor to extend the scope of previous investigations carried out by \cite{Mountrichas2021b, Mountrichas2022a, Mountrichas2022b} by comparing the SFR of AGN and non-AGN systems over a broader range of parameters, encompassing a wider span of L$_X$ ($\rm 42<log\,[L_{X,2-10keV}(ergs^{-1])}<45.5$) and M$_*$ ($\rm 9.5<log\,[M_*(M_\odot)] < 12.0$), with a specific emphasis on the lower M$_*$ regime. Secondly, we aim to revisit the SFR of type 1 and type 2 AGN while considering their L$_X$ and redshift dependencies.  For that purpose, we utilize the SFR$_{norm}$ parameter, defined as the ratio of the SFR of galaxies hosting AGN to the SFR of non-AGN systems, that share similar M$_*$ and redshift \citep[e.g.][]{Mullaney2015, Masoura2018}. The paper is organized as follows. In Sect. \ref{sec_data} we provide an overview of the parent catalogue used in our study. Sect. \ref{sec_analysis} elaborates on the SED fitting analysis and outlines the various criteria and requirements applied to select the final X-ray AGN and non-AGN galaxy samples. In Sect. \ref{sec_results}, we present the results of our analysis and in Sect. \ref{sec_conclusions} we summarize our main findings.

\section{Data}
\label{sec_data}

The parent catalogue used in our analysis, has been compiled within the framework of the project, "Athena: Scientific participation in the mission and development of the X-IFU instrument". To create this catalogue, the 10,242 fields from Data Release 8 (DR8) of the third XMM-Newton Serendipitous Source Catalogue (3XXM) were utilized. The aim was to identify sources in the optical, near-infrared (NIR), and mid-infrared (MIR) wavelength ranges that were included in the 3XMM footprint. To achieve this, we harnessed the data from the following surveys: the Dark Energy Survey \citep[DES;][]{Abbott2018}, the VISTA Hemisphere Survey \citep[VHS;][]{McMahon2013}, and the AllWISE survey \citep{Wright2010}. Among the 10,242 3XMM fields, a subset of 3,578 fields overlapped with VHS, and 1,674 of these fields also exhibited overlap with DES. This overlapping was defined by a radius of $15\arcmin$ measured from the center of the X-ray fields.

Upon obtaining the data from the aforementioned surveys, several data-cleaning steps were taken to ensure the quality and accuracy of the dataset. For instance, we excluded sky regions where an exceptionally bright source might obstruct the emission from other sources. Additionally, we identified and addressed cases of field overlap by grouping the central coordinates of X-ray fields based on their proximity. If two field centers were located within a distance of $30\arcmin$, it was indicative of field overlap. As a result of this overlap, a single source could appear multiple times in the initial tables. To resolve this, we eliminated these duplicate sources from the catalog, retaining only a single occurrence of each.

Finally, the cross-matching of the various tables was performed using the xmatch tool from the astromatch package\footnote{https://github.com/ruizca/astromatch}. This tool facilitated the matching of multiple catalogues and provided Bayesian probabilities for associations or non-associations, as detailed in \cite{Pineau2017, Ruiz2018}. We retained only those sources with a high probability of association, exceeding 68\% \citep[e.g.][]{Ruiz2018, Pouliasis2020}. In cases where one source was linked to multiple counterparts, we selected the association with the highest probability. We note that adjusting the probability threshold for association reduce the number of the sources used in our analysis (e.g., by 6\% and 14\% if the threshold is set to 80\% and 90\%, respectively), but it does not affect our overall results and conclusions.

The resulting catalog encompasses approximately 290,000 galaxies, all of which have detections in the DES, VHS, and AllWISE surveys. Within this sample, we establish the X-ray AGN dataset utilized in our analysis. Specifically, we focus on the 6,778 sources that are detected in X-rays and further narrow down the selection to those with $\rm log,[L_{X,2-10keV}(ergs^{-1})]>42$. L$_X$ is calculated using the X-ray fluxes available in the 3XMM catalogue. For the calculation, we assume a X-ray spectral index ($\Gamma$) of 1.7 \citep{Rosen2016} and we apply a conversion factor to scale the 4.5-12 keV, that is available in the 3XMM catalogue, to 2-10 keV, using the PIMMS website\footnote{https://cxc.harvard.edu/toolkit/pimms.jsp}. This criterion is met by 5,702 sources. The galaxies not detected in X-rays are utilized to identify sources for the control sample (as explained in the next section).


\section{Analysis}
\label{sec_analysis}

In this section, we outline the methodology employed to measure the host galaxy properties of the X-ray sources and describe the criteria utilized for the selection of sources with the most robust measurements and reliable classifications.

\subsection{Host galaxy properties}

The (host) galaxy properties of the X-ray AGN have been calculated via SED fitting, using the CIGALE code \citep{Boquien2019, Yang2020, Yang2022}. For consistency, we use the same models and parametric grid used in prior works that performed a similar analysis \citep[][]{Mountrichas2021b, Mountrichas2022a, Mountrichas2022b}.

In brief, the modeling of the galaxy component is accomplished through the use of a delayed Star Formation History (SFH) model with a functional form expressed as $\rm SFR\propto t \times \exp(-t/\tau)$. This model incorporates a star formation burst, as per \cite{Malek2018} and \cite{Buat2019}, as a continuous and consistent period of star formation spanning 50 million years (Myr). Stellar emission is described using the single stellar population templates sourced from \cite{Bruzual_Charlot2003} and is subject to attenuation following the attenuation law outlined by \cite{Charlot_Fall_2000}. For modeling nebular emission, CIGALE leverages the nebular templates rooted in the work of \cite{VillaVelez2021}. The emission stemming from dust heated by stars is accounted for in line with the approach introduced by \cite{Dale2014}, without any contribution from AGN sources. To incorporate AGN-related emission, CIGALE integrates the SKIRTOR models put forth by \cite{Stalevski2012} and \cite{Stalevski2016}. CIGALE has also the capability to model the X-ray emission of galaxies. In the SED fitting procedure, the X-ray flux in the $4.5-12$ keV energy band, as provided by the 3XMM catalogue is used. The parameter space used in the process of fitting SEDs can be found in Tables 1 within \cite{Mountrichas2021b, Mountrichas2022a, Mountrichas2022b}. The robustness and accuracy of the SFR measurements have been subject to thorough scrutiny in our prior research efforts, notably detailed in Section 3.2.2 of \cite{Mountrichas2022a}.

\subsection{Selection criteria and final samples} 
\label{sec_criteria}
Next, we describe the quality criteria and requirements we have applied to determine the sources eligible for inclusion in our final AGN and galaxy control samples.

\subsubsection{Criteria for SED fitting measurements}

In order to get reliable SED fitting results, it is essential to restrict the analysis to those sources with the highest possible photometric coverage. For that purpose, we require both the X-ray and the non-AGN galaxies in our datasets to have an extended photometric coverage. Specifically, following the works of \cite{Mountrichas2021b, Mountrichas2022a, Mountrichas2022b}, we require the sources to have available the following photometric bands $g, r, i, z, J, H, K$, W1, W2, W3 and W4. $g, r, i, z$ are the optical bands of DES, while $J, H, K$ and W1, W2, W3 and W4 are the photometric bands of VISTA and WISE, respectively. As previously mentioned, all 290\,000 sources meet this criterion.

Moreover, in alignment with previous studies, we implement stringent selection criteria to exclusively include sources with reliable SED fitting results. Specifically, we impose a reduced $\chi ^2$ threshold of $\chi ^2_r<5$ \citep[e.g.,][]{Masoura2018, Buat2021}. Furthermore, we exclude sources for which CIGALE was unable to effectively constrain the parameters of interest, namely SFR and M$_*$. CIGALE provides two values for each estimated galaxy property: one value corresponds to the best-fit model, while the other value (referred to as "bayes") represents the likelihood-weighted mean value. A significant disparity between these two calculations indicates a complex likelihood distribution and substantial uncertainties. Consequently, in our analysis, we only consider sources for which both $\rm \frac{1}{5}\leq \frac{SFR_{best}}{SFR_{bayes}} \leq 5$ and $\rm \frac{1}{5}\leq \frac{M_{*, best}}{M_{*, bayes}} \leq 5$ \citep[e.g.,][]{Buat2021, Koutoulidis2022, Pouliasis2022, Mountrichas2023a, Mountrichas2023, Mountrichas2023b, Mountrichas2023d}, where SFR$\rm _{best}$ and  M$\rm _{*, best}$ are the best-fit values of SFR and M$_*$, respectively and SFR$\rm _{bayes}$ and M$\rm _{*, bayes}$ are the Bayesian values estimated by CIGALE. 88\% and 77\% of the X-ray sources and the non-AGN galaxies meet these criteria, respectively.

Earlier research has established that absence of far-infrared photometry (e.g., Herschel) does not significantly affect the SFR calculations of CIGALE \citep{Mountrichas2021b, Mountrichas2022a, Mountrichas2022b}. At high redshifts (e.g., $\rm z>0.5$), the emission from young stars can be effectively traced using optical bands since the $u$ band shifts to rest-frame wavelengths of less than 2000\,$\AA$. However, at lower redshifts, it may be necessary to utilize shorter wavelengths to accurately capture the contribution of the young stellar population. \cite{Koutoulidis2022} demonstrated that the absence of both far-infrared and ultraviolet (UV) photometry does not compromise the reliability of CIGALE'S SFR calculations, particularly at low redshifts. Nonetheless, the photometric data from DES that we employ in our SED fitting analysis lacks information from the $u$ band. Consequently, to ensure the robustness of our analysis, we have included sources, encompassing both X-ray AGN and non-AGN, with redshifts exceeding $\rm z>0.3$. About 70\% of the X-ray sources and the non-AGN galaxies meet this requirement.  

\subsubsection{Exclusion of non-X-ray AGN systems from the galaxy control sample}

In order to make a meaningful comparison between the SFR of AGN and non-AGN samples, it is imperative not only to eliminate the 6\,778 X-ray-detected AGN from the galaxy control sample but also to exclude sources that might exhibit a substantial AGN contribution, which could potentially go undetected by X-ray observations \citep[e.g.,][]{Pouliasis2020}. To accomplish this, we rely on the measurements provided by CIGALE, specifically focusing on the AGN fraction parameter denoted as $\rm frac_{AGN}$. This parameter is defined as the ratio of AGN infrared emissions to the total infrared emissions of the galaxy, spanning the wavelength range of $1-1000\, \mu m$.

Consistent with the methodology employed in our earlier studies \citep{Mountrichas2021b, Mountrichas2022a, Mountrichas2022b}, we adopt a threshold that excludes sources with $\rm frac_{AGN}>0.2$ from the galaxy control sample. This criterion leads to the rejection of approximately 45\% of the galaxies within the reference sample. This percentage aligns with findings from our prior investigations. Furthermore, these studies have robustly demonstrated that the overall results and conclusions remain unaffected, regardless of whether these sources are included in the analysis or the choice of the threshold for $\rm frac_{AGN}$ \citep[see Section 3.3 in][]{Mountrichas2022a, Mountrichas2022b}.

\subsubsection{Mass completeness limits}

Calculation of the SFR$_{norm}$ parameter, requires both the AGN and the galaxy control samples to be mass complete within the redshift range of interest. For that purpose, similarly to our previous works, we use the method described in \cite{Pozzetti2010} to calculate the mass completeness limits of our datasets. Specifically, we use the galaxy control sample and the following expression that estimates the mass the galaxy would have if its apparent magnitude was equal to the limiting magnitude of the survey for a specific photometric band: 
\begin{equation}
\rm log\,M_{*,lim} = log M_*+0.4(m-m_{lim}).
\end{equation}
$\rm M_{*,lim}$ is the limiting M$_*$ of each galaxy at each redshift interval, M$_*$ is the stellar mass of each source measured by CIGALE, m is the AB magnitude of the source, and m$_{\rm lim}$ is the AB magnitude limit of the survey. We use $\rm K_s$ as the limiting band of the samples, in accordance with previous studies \citep{Laigle2016, Florez2020, Mountrichas2021c, Mountrichas2022a} and set $m_{lim}=23.06$ \citep{McMahon2013}. The process for the calculation of $\rm M_{*,lim}$ is described in detail in \cite{Mountrichas2021b, Mountrichas2022a, Mountrichas2022b}. We find that the stellar mass completeness limits of our galaxy reference catalogue is $\rm log\,[M_{*,95\%lim}(M_\odot)]= 9.61$, 10.23 and 10.98 at $\rm 0.3<z<1.0$, $\rm 1.0<z<2.0$ and $\rm 2.0<z<2.5$, respectively.

\subsubsection{Identification of quiescent systems}

Most previous studies that measured the SFR$_{norm}$ parameter, used analytical expressions from the literature to calculate the SFR of MS galaxies, with most commonly used the formulation presented in \cite{Schreiber2015} \citep{Mullaney2015, Masoura2018, Bernhard2019, Masoura2021, Pouliasis2022, Koutoulidis2022}. Hence, we recognize and exclude quiescent systems from our analysis, retaining only star-forming systems. Our goal is not to define our own MS, but to exclude in a uniform manner the majority of quiescent data from our samples. We note, though, that prior works have shown that the inclusion of quiescent systems in the analysis affects, mainly, the amplitude of the SFR$_{norm}$ measurements and not the observed trends \citep{Mountrichas2021b, Mountrichas2023b}. 

To discern quiescent systems, we adopt a methodology akin to that presented in detail in our previous works \citep{Mountrichas2021b, Mountrichas2022a, Mountrichas2022b}. This method relies on the calculation of the specific SFR ($\rm sSFR=\frac{SFR}{M_*}$) of each source. Specifically, we use the long tail or the position of the lower second peak present in the sSFR distributions, at different redshift intervals ($\rm 0.3<z\leq 1.0$, $\rm 1.0<z\leq 2.0$ and $\rm 2.0<z\leq 2.5$) to identify quiescent sources. About 10\% of the galaxies within the control sample are identified as quiescent. This number increases to $\sim 25\%$ for the X-ray AGN dataset. 

\begin{figure}
\centering
  \includegraphics[width=0.95\columnwidth, height=7.2cm]{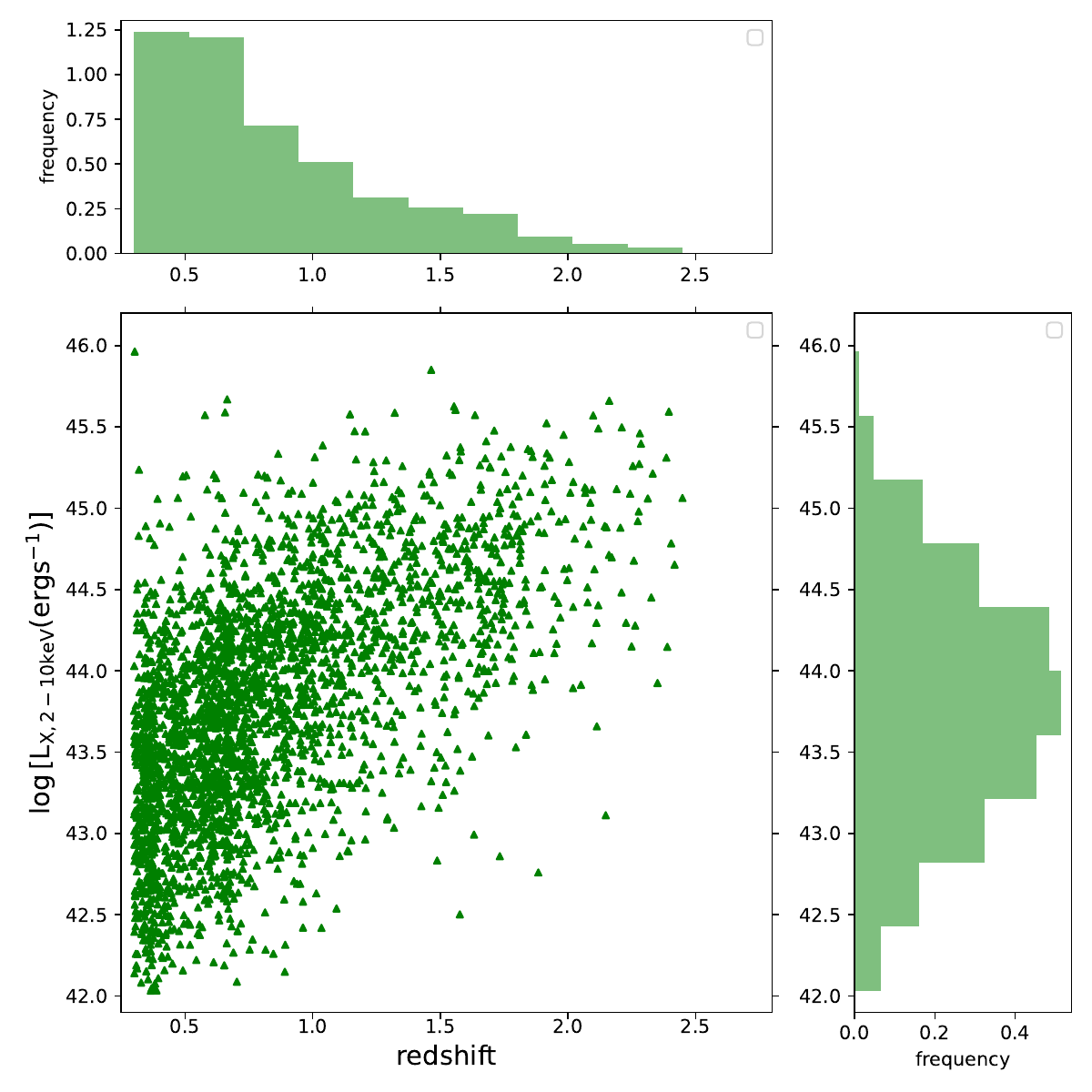}   
  \caption{Distribution in the (intrinsic) L$_X-$redshift plane, of the 2\,677 X-ray AGN used in our analysis.}
  \label{fig_lx_redz}
\end{figure}

Application of the all the criteria described above results in 2\,677 X-ray AGN and 64\,557 galaxies in the control sample (non-AGN), within a redshift range of $\rm 0.3<z<2.5$. Their distribution in the L$_X-$redshift plane is shown in Fig. \ref{fig_lx_redz}. These are the sources we use in the first part of our analysis (Sect. \ref{sec_sfrnorm_mstar}).

\begin{figure}
\centering
  \includegraphics[width=0.95\columnwidth, height=7.2cm]{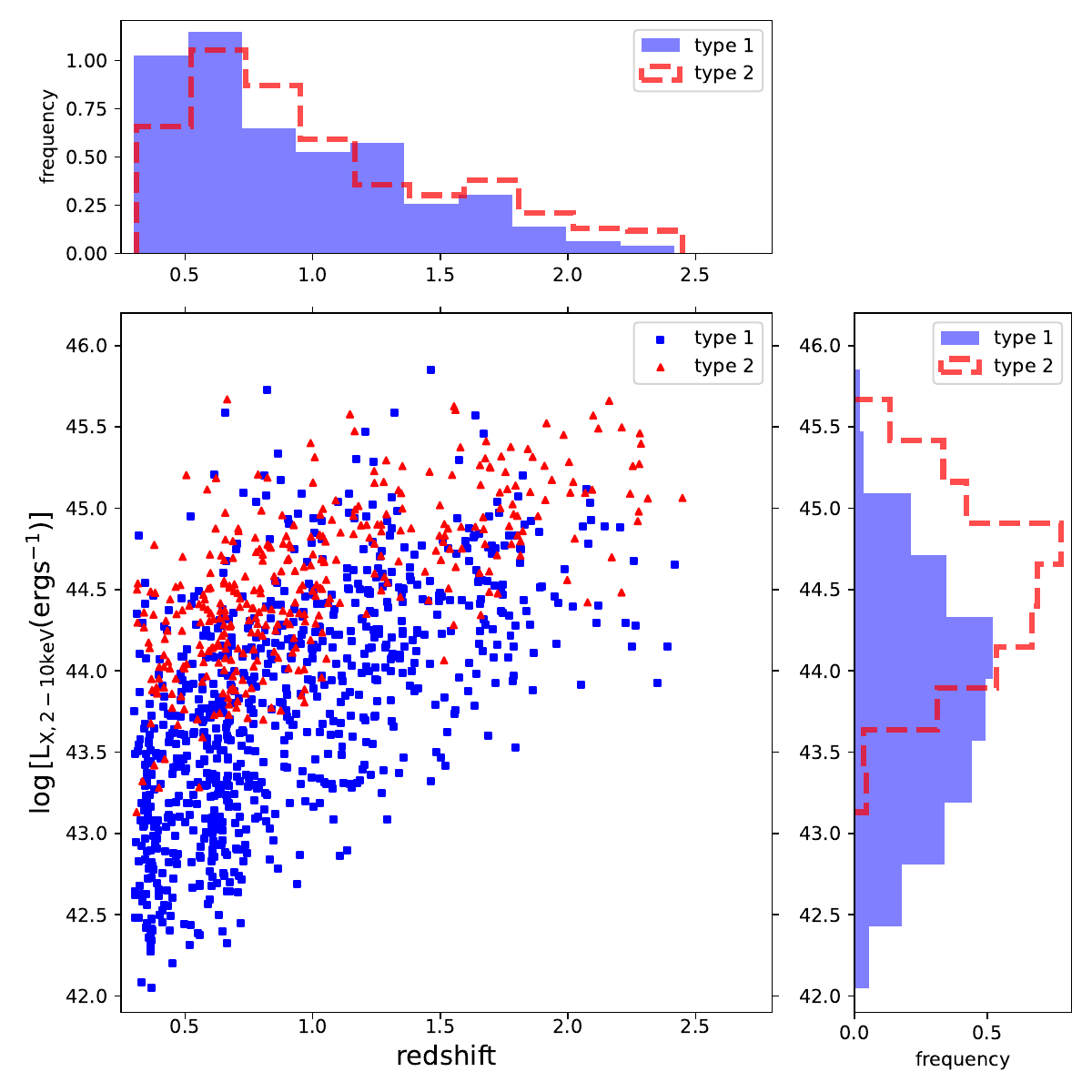}   
  \caption{Distribution in the L$_X-$redshift plane, of 825 type 1 (blue triangles) and 355 type 2 (red circles) AGN used in our analysis.}
  \label{fig_lx_redz_type}
\end{figure}  

\begin{figure*}
\centering
  \includegraphics[width=0.95\columnwidth, height=7.5cm]{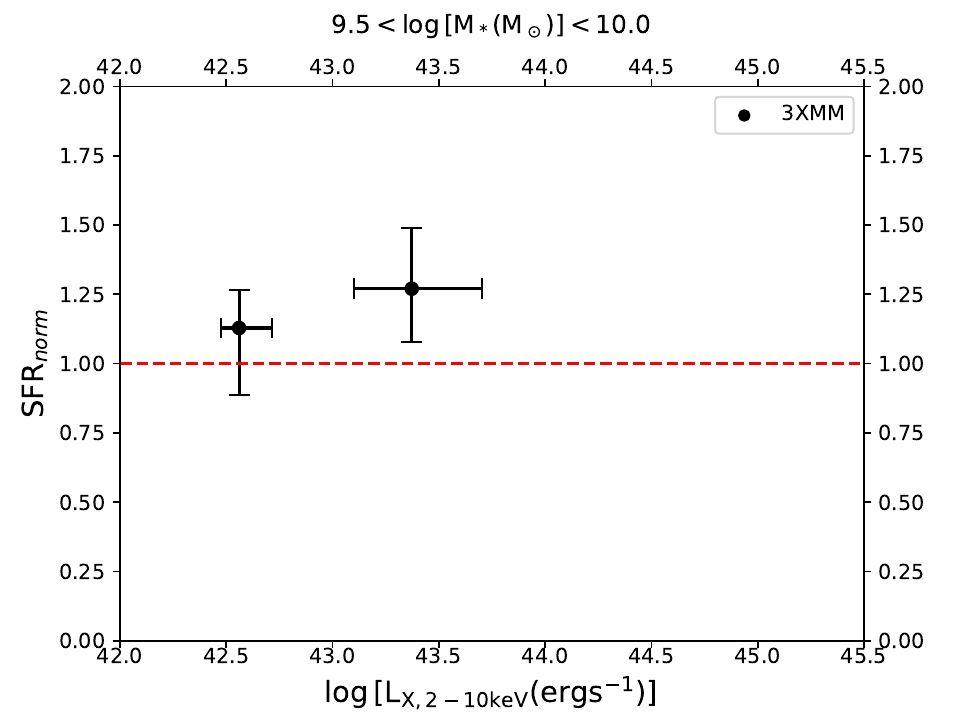}   
  \includegraphics[width=0.95\columnwidth, height=7.5cm]{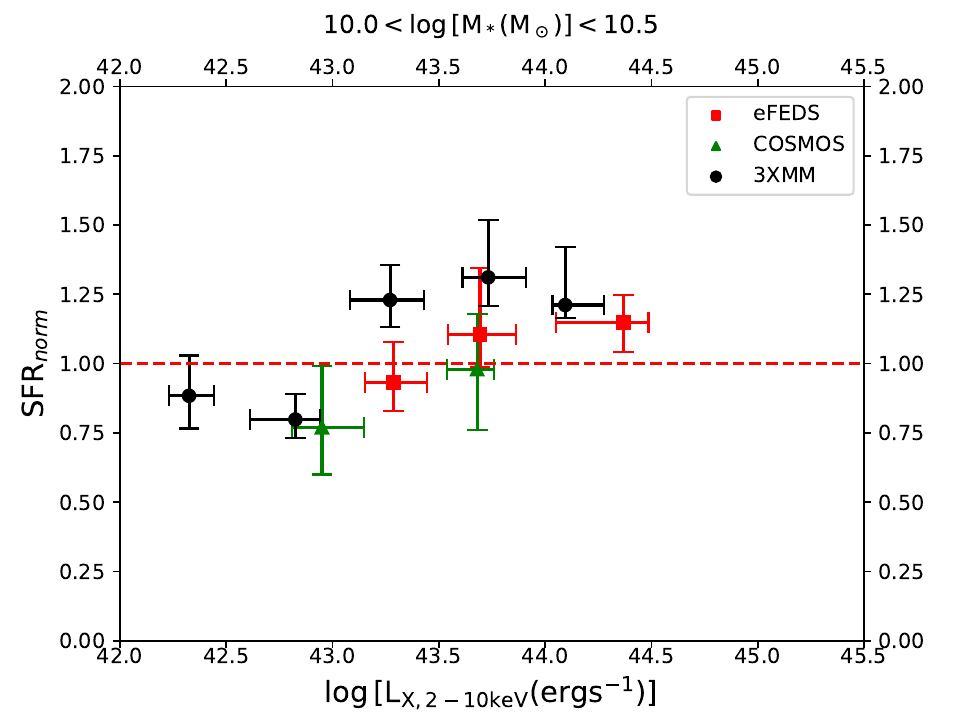} 
  \includegraphics[width=0.95\columnwidth, height=7.5cm]{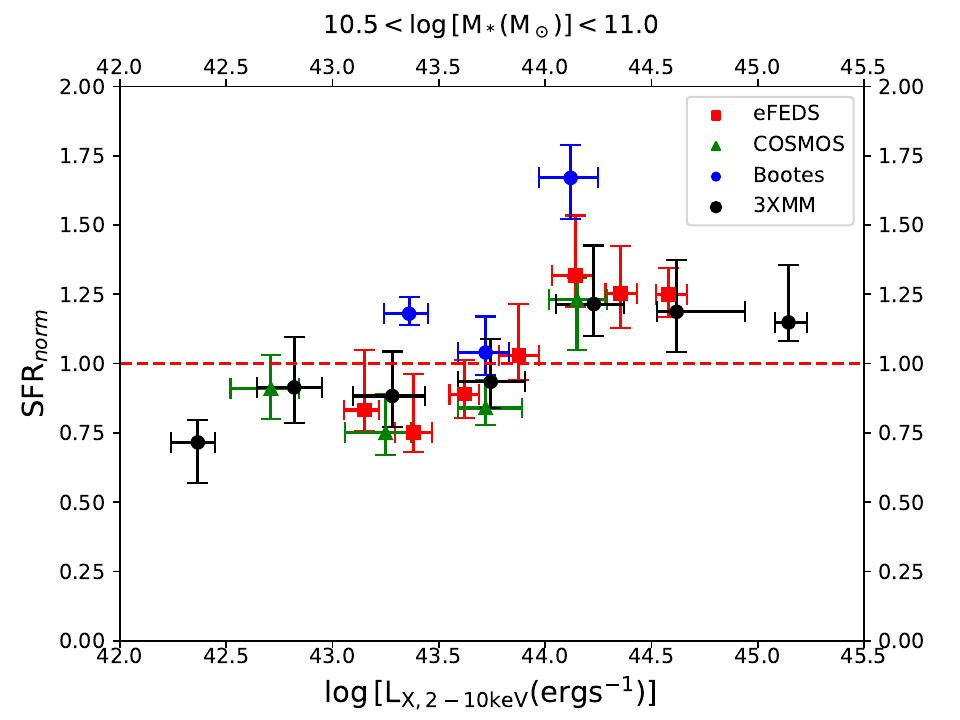}
  \includegraphics[width=0.95\columnwidth, height=7.5cm]{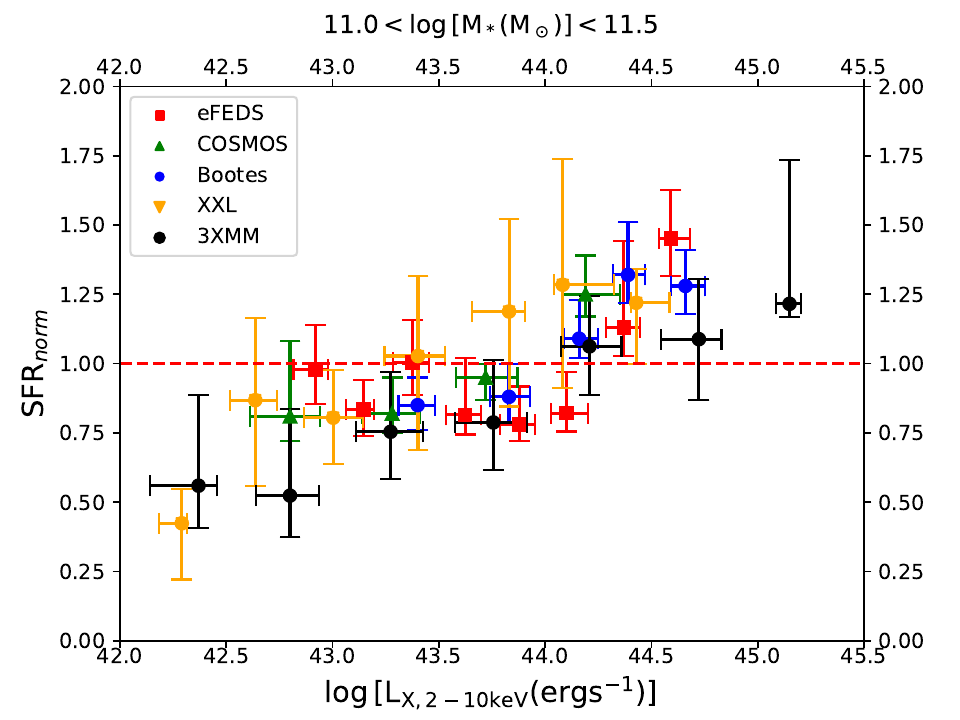}
  \includegraphics[width=0.95\columnwidth, height=7.5cm]{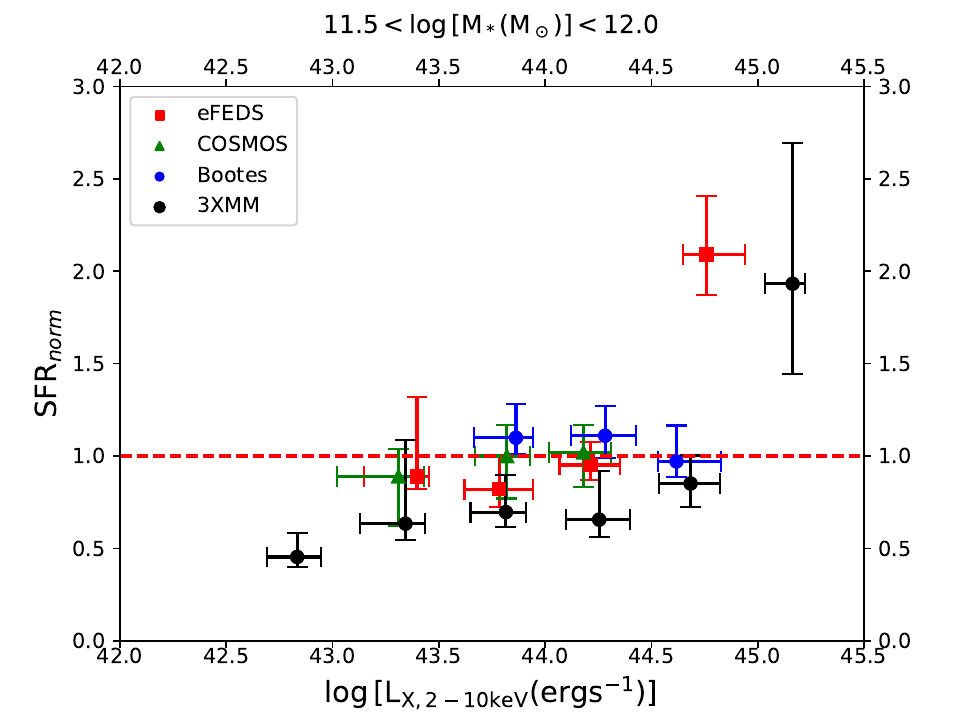}
  \caption{SFR$_{norm}$ vs. X-ray luminosity for five stellar-mass bins. We complement our results (black filled circles), with those using the Bo$\rm \ddot{o}$tes \citep{Mountrichas2021b}, the COSMOS \citep{Mountrichas2022a}, the eFEDS \citep{Mountrichas2022b} and the XMM-XXL \citep{Mountrichas2024d} datasets. The dashed horizontal line indicates the SFR$_{norm}$ value ($=1$) for which the SFR of AGN is equal to the SFR of non-AGN star-forming galaxies. The measurements are grouped in bins of L$_X$, of 0.5\,dex width, with the exception of the lowest M$_*$ range, where a L$_X$ bin size of 1\,dex has been chosen. Median values of SFR$_{norm}$ and L$_X$ are presented. Errors are calculated using bootstrap resampling.}
  \label{fig_sfrnorm_lx_all}
\end{figure*}

\begin{figure}
\centering
  \includegraphics[width=0.95\columnwidth, height=7.5cm]{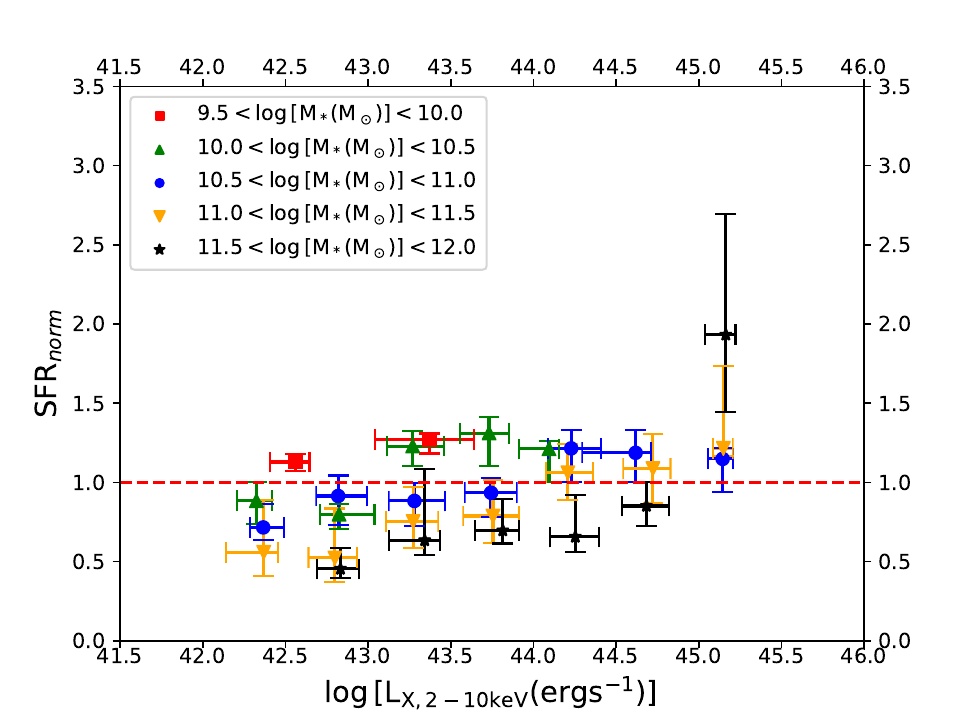}  
  \caption{Compilation of our SFR$_{norm}-$L$_X$ measurements, for different stellar masses, as indicated in the legend of the Figure.}
  \label{fig_sfrnorm_lx_mstar}
\end{figure}

\begin{figure}
\centering
  \includegraphics[width=1\columnwidth, height=7.5cm]{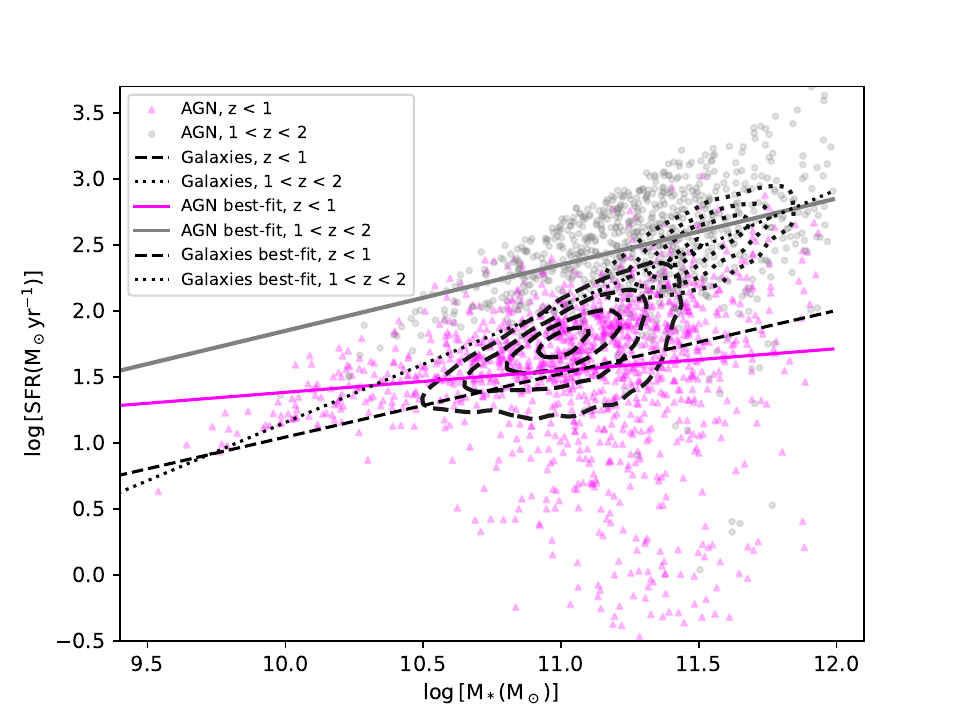}  
  \caption{SFR vs. M$_*$, for the X-ray AGN and non-AGN galaxies in our dataset, at different redshift intervals, as indicated in the legend. Lines show the best-fits for each subset.}
  \label{fig_sfr_mstar}
\end{figure}

\subsection{Classification of AGN}
\label{sect_classif}

To classify AGN into type 1 and 2 sources, we use the SED fitting measurements. Specifically, we employ the bayes and best estimates of the inclination, $i$, parameter, derived by CIGALE. We follow the criteria applied by \cite{Mountrichas2021b} and \cite{Mountrichas2024a} and classify as type 1 those with $i_{best}=30^{\circ}$ and $i_{bayes}<40^{\circ}$, while type 2 sources are those with $i_{best}=70^{\circ}$ and $i_{bayes}>60^{\circ}$.

In \cite{Mountrichas2021b}, CIGALE's classification was compared with the categorization provided in the catalog presented by \cite{Menzel2016}. In this catalog, AGN were divided into two categories: broad-line (type 1) and narrow-line (type 2) sources, based on the Full Width Half Maximum (FWHM) of emission lines originating from different regions of the AGN, including H$_\beta$, MgII, CIII and CIV. The analysis revealed that CIGALE exhibited an accuracy of approximately 85\% in classifying type 1 AGN. A similar level of accuracy was observed for the completeness of type 1 source identification. However, for type 2 sources, CIGALE's performance was approximately 50\%, both in terms of reliability and completeness. The reliability is defined as the fraction of the number of type 1 (or type 2) sources classified by the SED fitting that are similarly classified by optical spectra. The completeness refers to how many sources classified as type 1 (or type 2) based on optical spectroscopy were identified as such by the SED fitting results. For the purposes of our current study, our primary focus is on evaluating the reliability performance of CIGALE.

The reliability of approximately 85\% in CIGALE's identification of type 1 sources meets our acceptability criteria for the purposes of our statistical analysis. However, the reliability of the SED fitting code in the case of type 2 AGN is relatively lower, suggesting that roughly half of the sources classified as type 2 by CIGALE are indeed misclassified. Nonetheless, \cite{Mountrichas2021b} demonstrated that the majority ($\sim 82\%$) of these misclassified type 2 sources exhibit elevated polar dust values (E$_{B-V}>0.15$; refer to their Figure 8 and Section 5.1.1). Consequently, we have excluded these sources from our analysis and categorized as type 2 those AGN that meet the specified inclination angle criteria and also possess polar dust values lower than E$_{B-V}<0.15$ (i.e., similar to the type 2 classification criteria applied in \cite{Mountrichas2024a}. It is worth noting that the inclusion of polar dust in the fitting process enhances the accuracy of CIGALE's source type classification, particularly in terms of the reliability of identifying type 2 sources \citep[see Sect. 5.5 in][]{Mountrichas2021a}.

Application of these criteria on the 2\,677 AGN (see previous section) results in 825 type 1 and 355 type 2 AGN. Their L$_X-$redshift distribution is presented in Fig. \ref{fig_lx_redz_type}. These are the sources used in the second part of our analysis (Sect. \ref{sec_sfrnorm_type}).

\begin{figure}
\centering
  \includegraphics[width=0.95\columnwidth, height=7.5cm]{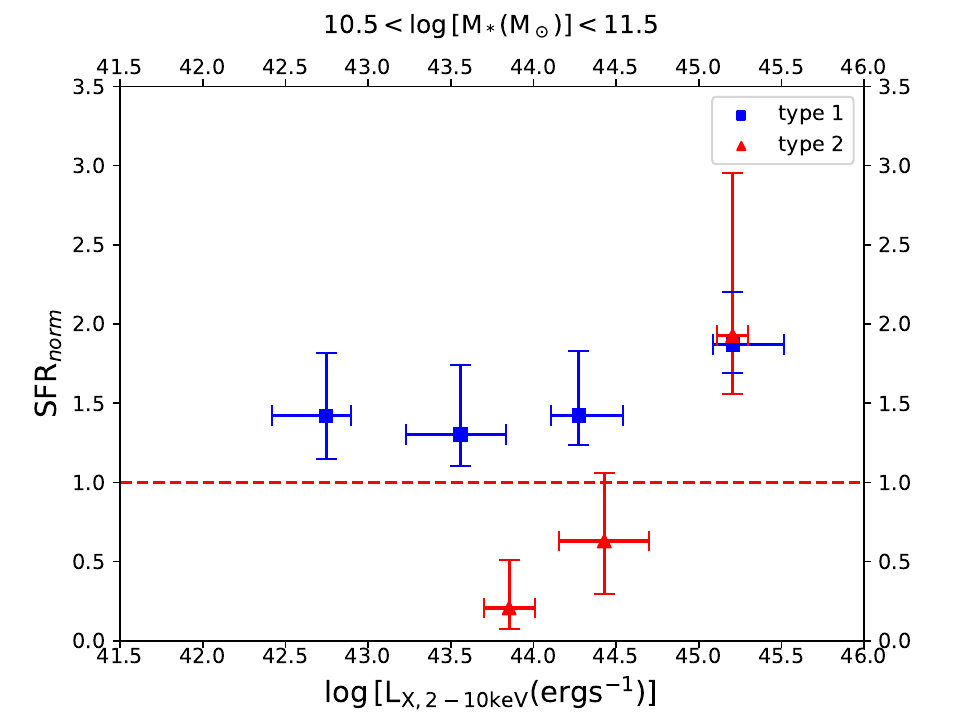}   
  \caption{SFR$_{norm}-$L$_X$ for different AGN types, at $\rm 0.3<z<2.5$.}
  \label{fig_sfrnorm_lx_type}
\end{figure}  

\begin{figure}
\centering
  \includegraphics[width=0.95\columnwidth, height=7.5cm]{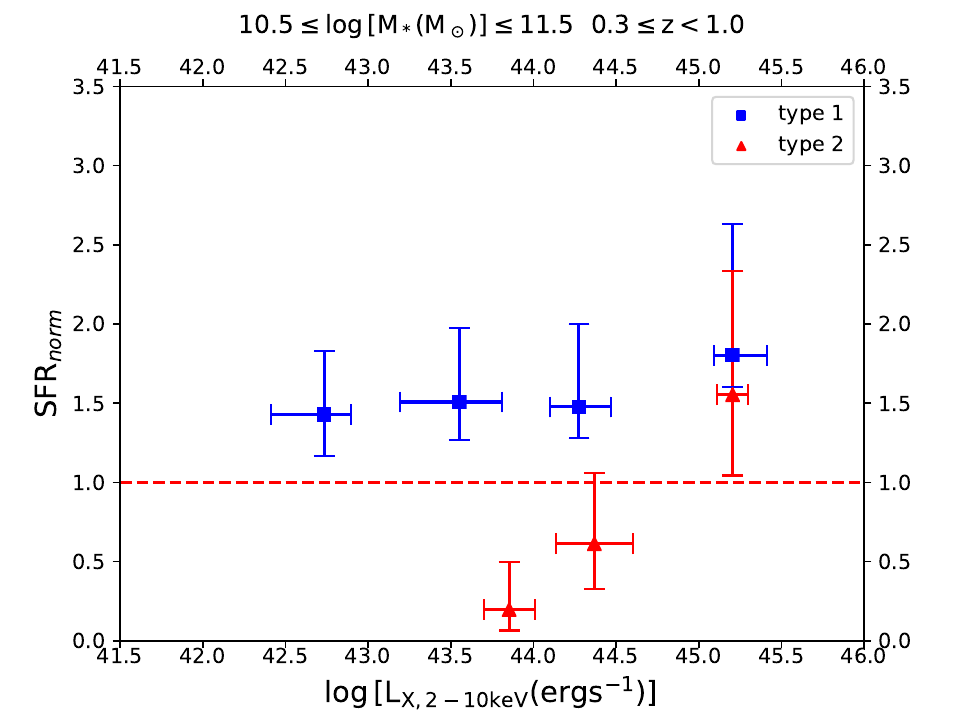}   
    \includegraphics[width=0.95\columnwidth, height=7.5cm]{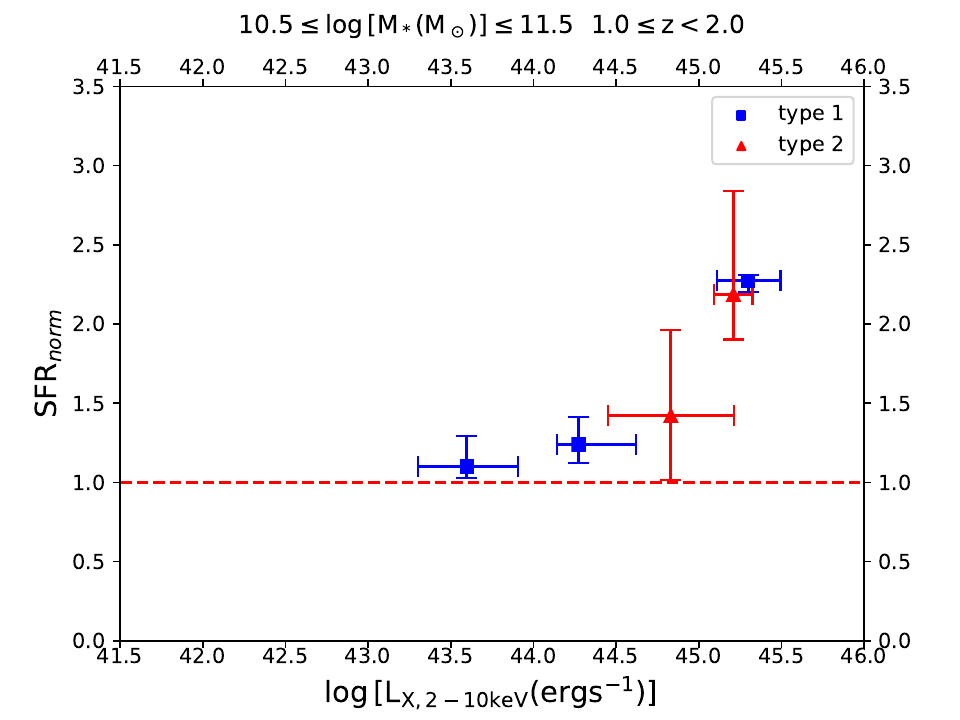}   
  \caption{SFR$_{norm}-$L$_X$ for type 1 (red triangles) and type 2 (blue squares) AGN. The top panel, presents the results at $\rm 0.3\leq z<1.0$. The bottom panel shows the measurements for sources within $\rm 1.0\leq z<2.0$.}
  \label{fig_sfrnorm_lx_type_redz}
\end{figure}

\section{Results}
\label{sec_results}

In this section, we present the results of our analysis. Specifically, we investigate the SFR$_{norm}-$L$_X$ relation, for galaxies of different M$_*$ and for different AGN types. 

\subsection{SFR$_{norm}$ as a function of L$_X$ and M$_*$}
\label{sec_sfrnorm_mstar}

To perform a comparison between the SFR of AGN and non-AGN galaxies, we follow the methodology presented, for instance, in \cite{Mountrichas2021b, Mountrichas2022a, Mountrichas2022b, Mountrichas2023b, Mountrichas2023d}. Specifically, we employ the SFR$_{norm}$ parameter. For the calculation of SFR$_{norm}$, we utilize the galaxy control sample presented in Sect. \ref{sec_analysis}. Utilizing a galaxy reference catalogue minimizes systematic effects that may affect the accuracy of the SFR$_{norm}$ calculation, compared to using analytical expressions from the literature \citep[e.g.,][]{Schreiber2015} for the calculation of the SFR of non-AGN galaxies \citep{Mountrichas2021b}. 

To measure the SFR$_{norm}$ parameter,  we divide the SFR of each X-ray AGN by the SFR of galaxies in the control sample that closely match the AGN in terms M$_*$ within $\pm 0.2$ dex, redshift within $\pm 0.075 \times (1+z)$. Furthermore, each source's contribution is weighted based on the uncertainties associated with the SFR and M$_*$ measurements obtained using the CIGALE methodology. The median values of these ratios are subsequently utilized as the SFR$_{norm}$ for each X-ray AGN. It's worth noting that our measurements are not significantly affected by the specific size of the region surrounding the AGN. However, selecting smaller regions does have an impact on the accuracy of the calculations, as discussed in \cite{Mountrichas2021b}.

The results of our measurements are presented in Fig. \ref{fig_sfrnorm_lx_all}. Each panel corresponds to systems with different M$_*$ range. Previous studies have shown that there is no (strong) evolution of the SFR$_{norm}-$L$_X$ relation with redshift \citep{Mountrichas2021b, Mountrichas2022a, Mountrichas2022b}. Therefore, we do not split our measurements into different redshift intervals. Median values of SFR$_{norm}$ and L$_X$ are presented. The bins are grouped in bins of L$_X$, of 0.5\,dex width, with the exception of the lowest M$_*$ range, where a L$_X$ bin size of 1\,dex has been chosen, due to the low number of X-ray sources with  $\rm 9.5<log\,[M_*(M_\odot)]<10.0$. The errors presented are $1\,\sigma$, calculated using bootstrap resampling. Only bins that include ten or more sources are presented in the plots. We have overlaid our results with those presented in Fig. 5 in \cite{Mountrichas2022b}, where they amalgamated findings from a similar analysis based on data in the Bo$\rm \ddot{o}$tes, COSMOS and eFEDS fields. Furthermore, we have incorporated data from a study conducted by \cite{Mountrichas2024d}, which employed X-ray AGN data from the XMM-XXL field. It is important to highlight that while the latter study investigated the SFR$_{norm}-$L$_X$ relationship for systems falling within the range of $\rm 10.5 < \log\,[M_*(M_\odot)] < 11.5$, we have opted to compare their results with our findings within the range of $11.0 < \log\,[M_*(M_\odot)] < 11.5$. Additionally, it is worth noting that their results do not include the exclusion of quiescent systems from the X-ray and galaxy control samples.

For systems with intermediate M$_*$, that is $\rm 10.5<log\,[M_*(M_\odot)]<11.0$ and $\rm 11.0<log\,[M_*(M_\odot)]<11.5$, presented in the left and right panels of the middle row of Fig. \ref{fig_sfrnorm_lx_all}, we confirm the results of prior studies. Specifically, we find that AGN with low-to-intermediate L$_X$ ($\rm log\,[L_{X,2-10keV}(ergs^{-1})]<44$) present lower or at most equal SFR with that of non-AGN galaxies (SFR$_{norm}\leq 1$), while more luminous AGN have enhanced SFR by 20-30\% compared to galaxies without an AGN. 

More importantly, we observe that for the most massive systems (bottom panel in Fig. \ref{fig_sfrnorm_lx_all}), the SFR$_{norm}-$L$_X$ relation remains relatively constant up to an L$_X$ threshold,mirroring the observed trend in systems with $\rm 10.5<log\,[M_*(M_\odot)]<11.5$. However, the position of this threshold is higher in the case of these massive systems, at $\rm log,[L_{X,2-10keV}(ergs^{-1})]=45$. Beyond this threshold, we detect a substantial increase, roughly by a factor of two, in the SFR of galaxies hosting AGN in comparison to those without AGN. Compared to previous studies, our findings may seem slightly lower, however, they are statistically consistent with those earlier results. Notably, our findings reaffirm previous observations of a substantial elevation in SFR$_{norm}$ at very high L$_X$. Specifically, in a study by \cite{Mountrichas2022b}, data derived from the eFEDS field, incorporating X-ray observations from the eROSITA satellite, revealed a significant rise in SFR$_{norm}$ at an L$_X$ of approximately $\rm log\,[L_{X,2-10keV}(ergs^{-1})] \approx 45$. However, their dataset did not span higher L$_X$ values to ascertain whether this result was consistent or merely a statistical fluctuation. Our measurements validate the notion that, in the most massive systems, galaxies with AGN exhibit heightened SFRs in comparison to non-AGN galaxies, however, this enhancement is only observed at very high L$_X$. 

The outcomes pertaining to the least massive systems within our datasets are depicted in the upper panels of Figure \ref{fig_sfrnorm_lx_all}. In the case of galaxies falling within the range of $\rm 10.0 < log\,[M_*(M_\odot)] < 10.5$, we observe an augmentation in the SFR of AGN in comparison to non-AGN galaxies (indicated by SFR$_{norm} > 1$). This phenomenon mirrors what is seen in systems with intermediate stellar mass (i.e., $10.5 < \log\,[M_*(M_\odot)] < 11.5$). However, it is worth noting that this increase in the SFR$_{norm}$ parameter is detected at lower values of L$_X$ (i.e., approximately $\rm log\,[L_{X,2-10keV}(ergs^{-1})]\sim 43-43.5$), as opposed to the intermediate stellar mass galaxies where it occurs at around $\rm log\,[L_{X,2-10keV}(ergs^{-1})]\sim 44$.

Our results are in agreement with prior studies that have reported either a lower or similar SFR between low-to-moderate L$_X$ AGN and non-AGN galaxies \citep[e.g.,][]{Shimizu2015, Shimizu2017, Masoura2018, Bernhard2019} and an enhanced SFR of luminous AGN compared to their non-AGN counterparts \citep[e.g.][]{Florez2020, Pouliasis2022}. Our findings also underline the importance of M$_*$ in the comparison of the SFR of the two populations, similar to the results from recent studies \citep[e.g.][]{Torbaniuk2021, Torbaniuk2023}.

Overall, our findings underscore that the assessment of SFR in AGN-hosting and non-AGN galaxies hinges on both the power of AGN (L$_X$) and the M$_*$ of the hosting galaxy. Our results suggest that galaxies with AGN tend to exhibit elevated SFR when contrasted with those lacking AGN, after an L$_X$ threshold. However, the point at which the AGN start to present enhanced SFR compared to non-AGN (i.e., SFR$_{norm}>1$) varies depending on the M$_*$ of the host galaxy. More precisely, the threshold L$_X$ value for this enhancement increases as we transition to more massive galactic systems. These findings align with a hypothesis wherein AGN feedback, potentially manifested as strong winds \citep[e.g.,][]{Debuhr2012}, could lead to the overcompression of existing cold gas within the host galaxy \citep[e.g.][]{Zubovas2013}, consequently promoting star formation (positive feedback). The more massive the host galaxy, the stronger (more luminous) the AGN needs to be in order to exert an impact on the star formation within its host.

Lastly, Fig. \ref{fig_sfrnorm_lx_mstar} consolidates all our findings across different ranges of M$_*$. We observe that, for the same L$_X$, the amplitude of the SFR$_{norm}$ parameter diminishes as we transition to more massive galaxies, at least up to $\rm log\,[L_{X,2-10keV}(ergs^{-1})]=45$. This could be attributed to the fact that, in galaxies that host AGN the increase in SFR with rising M$_*$ is not as pronounced as it is in galaxies devoid of AGN. This variation could be a result of some of the available gas being channeled towards fueling the SMBH instead. This scenario implies that the X-ray main sequence \citep{Aird2018} has a less steep slope compared to the galaxy main sequence. Fig. \ref{fig_sfr_mstar} presents the SFR vs. M$_*$ for X-ray AGN and non-AGN galaxies, at different redshift intervals, as indicated in the legend of the plot. Within the redshift range of $\rm 0.3<z<1.0$, we observe a slope of $0.21\pm 0.04$ for AGN and $0.48\pm 0.01$ for non-AGN galaxies. In the redshift range of $\rm 1.0<z<2.0$, AGN exhibit a slope of $0.50\pm 0.04$, while galaxies without AGN display a slope of $0.88\pm 0.02$. Utilizing mean square error (MSE) analysis demonstrates an excellent goodness of fit for all the models (MSE value $<0.5$ in all instances). Additionally, consistent fits are achieved when employing the linmix module \citep{Kelly2007}, which conducts linear regression between two parameters by iteratively perturbing the data points within their uncertainties. These findings reinforce the interpretation mentioned earlier.

\subsection{SFR$_{norm}-$L$_X$ for type 1 and 2 AGN}
\label{sec_sfrnorm_type}

In this section, we compare the SFR$_{norm}-$L$_X$ relation for different AGN types. For that purpose, we classify the X-ray sources into type 1 and 2 AGN, using the results of CIGALE, as described in Sect. \ref{sect_classif}. Additionally, we narrow down our selection to sources falling within the M$_*$ range of $\rm 10.5<log\,[M_*(M_\odot)]<11.5$. We make this choice because, as both our study and previous research have indicated, the SFR$_{norm}$-L$_X$ relationship exhibits similarities within this M$_*$ range. This filtering reduces the number of AGN available for our analysis to 652 type 1 AGN and 293 type 2 AGN. 

The results are presented in Fig. \ref{fig_sfrnorm_lx_type}. Measurements are grouped in L$_X$ bins of size 1\,dex. As previously mentioned, we only present bins that include $\geq 10$ sources. We notice that type 1 AGN have higher SFR$_{norm}$ values compared to type 2, at least for AGN with L$_X$ within the range of $\rm 43.5<log\,[L_{X,2-10keV}(ergs^{-1})]<45$. Furthermore, within this L$_X$ range, we observe that type 1 AGN appear to have higher SFR compared to non-AGN galaxies of similar M$_*$ and redshift (i.e., SFR$_{norm}>1$). On the contrary, type 2 AGN tend to have lower SFR compared to galaxies without AGN. 

\cite{Masoura2021} conducted a study involving more than 3\,000 X-ray AGN in the XMM-XXL field, focusing on the SFR$_{norm}$-L$_X$ relationship for X-ray obscured and unobscured sources. Their classification criterion was based on the hydrogen column density, N$_H$. Specifically, they categorized sources with N$_H > 10^{21.5}$\,$\rm cm^{-2}$ as absorbed sources. According to their findings, they did not identify significant distinctions in the SFR$_{norm}$ as a function of L$_X$ between the two AGN categories. We note, though, that as several previous studies have emphasized, the adoption of different criteria for characterizing AGN based on their level of obscuration can lead to varying categorizations of AGN \citep[e.g.,][]{Merloni2014, Li2019, Masoura2020, Mountrichas2021b}. 


Some prior studies that compared the SFR of type 1 and 2 AGN, using optical spectra for the classification, did not discover significant differences in the SFR of the two AGN populations \citep{Zou2019, Mountrichas2021b}. However, it is crucial to emphasize that the AGN in these studies covered, mainly, lower L$_X$ values ($\rm log\,[L_{X,2-10keV}(ergs^{-1})]<44$) compared to our sources. Earlier research has underlined that the comparison of the host galaxy properties of different AGN types depends on the L$_X$ regime under consideration \citep{Georgantopoulos2023}). \cite{Mountrichas2024a}, employed X-ray sources in the eFEDS and COSMOS fields and classified them into type 1 and 2, using CIGALE's classification measurements, akin to our approach. According to their findings, the comparison of the SFR of type 1 and 2 AGN depends on both redshift and L$_X$. Specifically, they found that type 1 AGN tend to have higher SFR compared to type 2 sources, for sources with $\rm log\,[L_{X,2-10keV}(ergs^{-1})]<44$,  at all redshifts spanned by their dataset ($\rm 0.5<z<3.5$). Based on their results, this picture reverses at $\rm z>2$ and $\rm log\,[L_{X,2-10keV}(ergs^{-1})]>44$. At intermediate redshift ranges ($\rm 1<z<2$) and for $\rm log\,[L_{X,2-10keV}(ergs^{-1})]>44$ the two AGN populations appeared to have similar SFR.

In light of these findings, we divide our AGN sample into two redshift intervals, that is $\rm 0.3\leq z\leq 1.0$ and $\rm 1.0\leq z\leq 2.0$. At $\rm z>2$, we lack a sufficient number of type 2 sources to perform a meaningful analysis. The results are presented in Fig. \ref{fig_sfrnorm_lx_type_redz}. Notably, in the lower redshift interval (top panel), type 1 AGN appear to have higher SFR$_{norm}$ values compared to type 2, while at the higher redshift range (bottom panel), the two AGN types exhibit consistent SFR$_{norm}$ measurements. Although, the number of available sources for this exercise, and in particular the number of type 2 AGN, is not sufficiently large for strong conclusions to be drawn, these results appear to be in line with those presented in \cite{Mountrichas2024a}. Moreover, for type 2 AGN, SFR$_{norm}$ increases with L$_X$ at both redshift intervals, while for type 1, it remains roughly constant at low redshift and increases with L$_X$ at $\rm z>1$. 

The observed SFR$_{norm}>1$ for type 1 AGN in the specified L$_X$ range, suggests a potential positive feedback mechanism, where the presence of type 1 AGN enhances star formation beyond what is typical for galaxies without AGN. Type 1 AGN typically exhibit a clearer view of the central engine due to the absence of significant obscuration. This unimpeded view may lead to a more direct interaction between the AGN and the surrounding gas, potentially influencing star formation. The distinct behavior of type 1 AGN, where SFR$_{norm}$ remains relatively constant at low redshift and increases with L$_X$ at higher redshifts, might indicate that the feedback mechanisms associated with type 1 AGN evolve differently over cosmic time. The lower SFR$_{norm}$ in Type 2 AGN, at least at $\rm z<1$, may imply that the gas reservoirs in type 2 AGN host galaxies are less conducive to star formation, potentially due to feedback effects from the AGN. However, larger samples are required for strong conclusions to be drawn.

\section{Conclusions}
\label{sec_conclusions}

In this work, we compiled a dataset comprising of 2\,677 X-ray AGN detected by the XMM satellite, along with a control sample of 64\,557 galaxies without AGN, all of which lie in the 3XMM footprint. We have constructed SEDs for these sources by using photometric data from the DES, VHS and AllWISE surveys and employed the CIGALE SED fitting code to obtain measurements for their (host) galaxy properties. Our sources  span a wide parameter space, with objects falling within the ranges of $\rm 9.5<log\,[M_*(M_\odot)]<12.0$, $\rm 42<log\,[L_{X,2-10keV}(ergs^{-1})]<45.5$ and $\rm 0.3<z<2.5$. Leveraging CIGALE's measurements, we classified AGN into type 1 and 2. In our analysis, we have used 652 type 1 AGN and 293 type 2 within a M$_*$ range of $\rm 10.5<log\,[M_*(M_\odot)]<11.5$. The main results of our investigation can be summarized as follows:

\begin{itemize} 

\item[$\bullet$] The comparison of the SFR of AGN and non-AGN galaxies hinges on both the L$_X$ and the stellar mass. Specifically, AGN tend to present enhanced SFR when compared to non-AGN systems, but the L$_X$ at which this elevation becomes apparent increases as we transition to more massive galactic systems.  

\item[$\bullet$] For the same L$_X$, the amplitude of the SFR$_{norm}$ parameter decreases as we move to more massive galaxies. This could be attributed to the fact that, in AGN the increase in SFR with rising M$_*$ is not as evident as it is in galaxies without AGN. This scenario is supported by the less steep slope we observe for the X-ray star-forming main sequence compared to the galaxy main sequence.

\item[$\bullet$] For systems with $\rm 10.5<log\,[M_*(M_\odot)]<11.5$, type 1 AGN tend to exhibit higher SFR compared to type 2, at matching L$_X$ and M$_*$, at $\rm z<1$. However, at higher redshifts the two AGN populations present similar SFR.

\item[$\bullet$] At low redshifts ($\rm z<1$) and  $\rm 10.5<log\,[M_*(M_\odot)]<11.5$, type 1 AGN have enhanced SFR compared to non-AGN systems, with similar M$_*$ and redshift. On the contrary, type 2 AGN have lower SFR compared to galaxies without AGN, at least up to $\rm log\,[L_{X,2-10keV}(ergs^{-1})]<45$. 

\item[$\bullet$] At higher redshifts ($\rm z>1$) and  $\rm 10.5<log\,[M_*(M_\odot)]<11.5$, both type 1 and type 2 AGN tend to have higher SFR than non-AGN galaxies with similar redshift and M$_*$. 

\end{itemize}


\begin{acknowledgements}
This project has received funding from the European Union's Horizon 2020 research and innovation program under grant agreement no. 101004168, the XMM2ATHENA project. VAM acknowledges support by the Grant RTI2018-096686-B-C21 funded by MCIN/AEI/10.13039/501100011033 and by ’ERDF A way of making Europe’. AC and FJC acknowledge support by the Grant PID2021-122955OB-C41 funded by MCIN/AEI/10.13039/501100011033 and by ERDF A way of making Europe. This research has made use of TOPCAT version 4.8 \citep{Taylor2005} and Astropy \citep{PriceWhelan2022}.

\end{acknowledgements}

\bibliography{mybib}
\bibliographystyle{aa}

\end{document}